\documentclass[11pt]{article}
\usepackage{jheppub}
\pdfoutput=1
\usepackage[bottom]{footmisc}
\usepackage{amsfonts}
\usepackage{graphicx}
\usepackage{amsmath}
\usepackage{amssymb}
\usepackage{mathrsfs}

\usepackage[normalem]{ulem}
\usepackage[utf8]{inputenc}

\usepackage[english]{babel}
\usepackage{indentfirst}
\usepackage[dvipsnames]{xcolor}

\usepackage{graphicx}
\usepackage{tikz-cd}
\usepackage{xfrac}
\usepackage{url}
\usepackage{color}

\numberwithin{equation}{section}
\date{\today}

\def\be{\begin{equation}}
\def\ee{\end{equation}}
\def\ba#1\ea{\begin{align}#1\end{align}}

\def\la{\langle}
\def\ra{\rangle}

\newcommand{\bb}[1]{\mathbb{#1}}

\newcommand{\scr}[1]{\mathscr{#1}}

\newcommand{\wt}[1]{\widetilde{#1}}

\newcommand{\wh}[1]{\widehat{#1}}

\newcommand{\gb}[1]{[\hspace{-0.13em}[#1]\hspace{-0.13em}]}

\title{Guiding center quantization of a\\ quantum Hall analog of Hawking radiation} 
\author[a]{Rodrigo Andrade e Silva}
\author[b]{and Ted Jacobson}
\affiliation[a]{Perimeter Institute for Theoretical Physics, Waterloo, Canada}
\affiliation[b]{Maryland Center for Fundamental Physics,
University of Maryland, College Park, MD, USA}
\emailAdd{randradeesilva@perimeterinstitute.ca}
\emailAdd{jacobson@umd.edu}
\date{}

\abstract{We revisit the quantum Hall analog of Hawking radiation 
in which a Fermi sea of electrons occupying half of a plane, 
subjected to a quadrupolar electric potential, gives rise to 
analog Hawking radiation 
of chiral edge modes.
We show that the phenomenon is accurately captured by quantization of the  classical 
guiding center theory, where the gyroscopic motion of the electrons is 
coarse grained and only the drift motion is resolved. 
The quantum dynamics takes place on
a non-commutative plane,
which requires an electron localized in the half-plane 
to be supported everywhere along the edge direction.
This kinematical constraint on the quantum state leads directly to the
radiation, which propagates in both directions away from the origin
along the edge. The radiation is thermal with respect to laboratory time,
which is equal (up to a constant factor) to
the boost angle in the analog Minkowski spacetime
in which the chiral edge modes propagate, so in fact it corresponds to 
analog Unruh radiation.}

\begin{document}

\maketitle
\flushbottom

\vspace{\baselineskip}

\vspace{10pt}
\noindent\rule{\textwidth}{1pt}
\tableofcontents
\noindent\rule{\textwidth}{1pt}
\vspace{10pt}

\section{Introduction}
\label{sec:intro}

Numerous analog models of Hawking radiation and (fewer) of the Unruh
effect have been devised and investigated\cite{Barcelo:2026vlr},
motivated by the difficulty of observing these effects in their original spacetime setting, 
by the guidance they might provide for the transPlanckian problem \cite{tHooft:1984kcu, tHooft:1994tah} 
and the origin of the outgoing black hole modes \cite{Unruh:1980cg, Jacobson:1999zk},
and by the promise that identifying and observing analogous phenomena in condensed matter systems may reveal interesting behavior of those systems in their own right. 
In this paper we study aspects of one such analog, 
in which the role of the quantum field is played by the chiral edge modes
of a quantum Hall system. This model was introduced by M.~Stone\cite{Stone:2012cx},
and was later further studied in \cite{Hegde:2018xub, Subramanyan:2020fmx}.

We have two aims here: one is  
to clarify how the chiral nature of the edge modes, and the
proportionality of inertial laboratory time to the boost angle of the Rindler wedge in
an analog Minkowski spacetime, allow the
model to serve as both an Unruh effect analog and a Hawking effect analog;
the other is to demonstrate how the model can be analyzed 
within the effective theory known as the guiding center approximation,
in which the gyro motion of the electrons is not resolved. To carry out this 
analysis, we canonically quantize the classical guiding center theory,
in which the phase space coincides with the spatial plane, which thus becomes
a noncommutative space. (In another paper \cite{AndradeeSilva:2026pxp} 
we study in detail several aspects 
of the quantized guiding center theory, and its relation to the microscopic theory.)
We find that the Unruh/Hawking effect  is faithfully captured in 
the guiding center description. 

\section{Hall effect edge mode Hawking radiation}
\label{sec:StoneModel}
\begin{figure}[ht!]  
\centering
\hspace*{-1cm}\includegraphics[scale = 0.25]{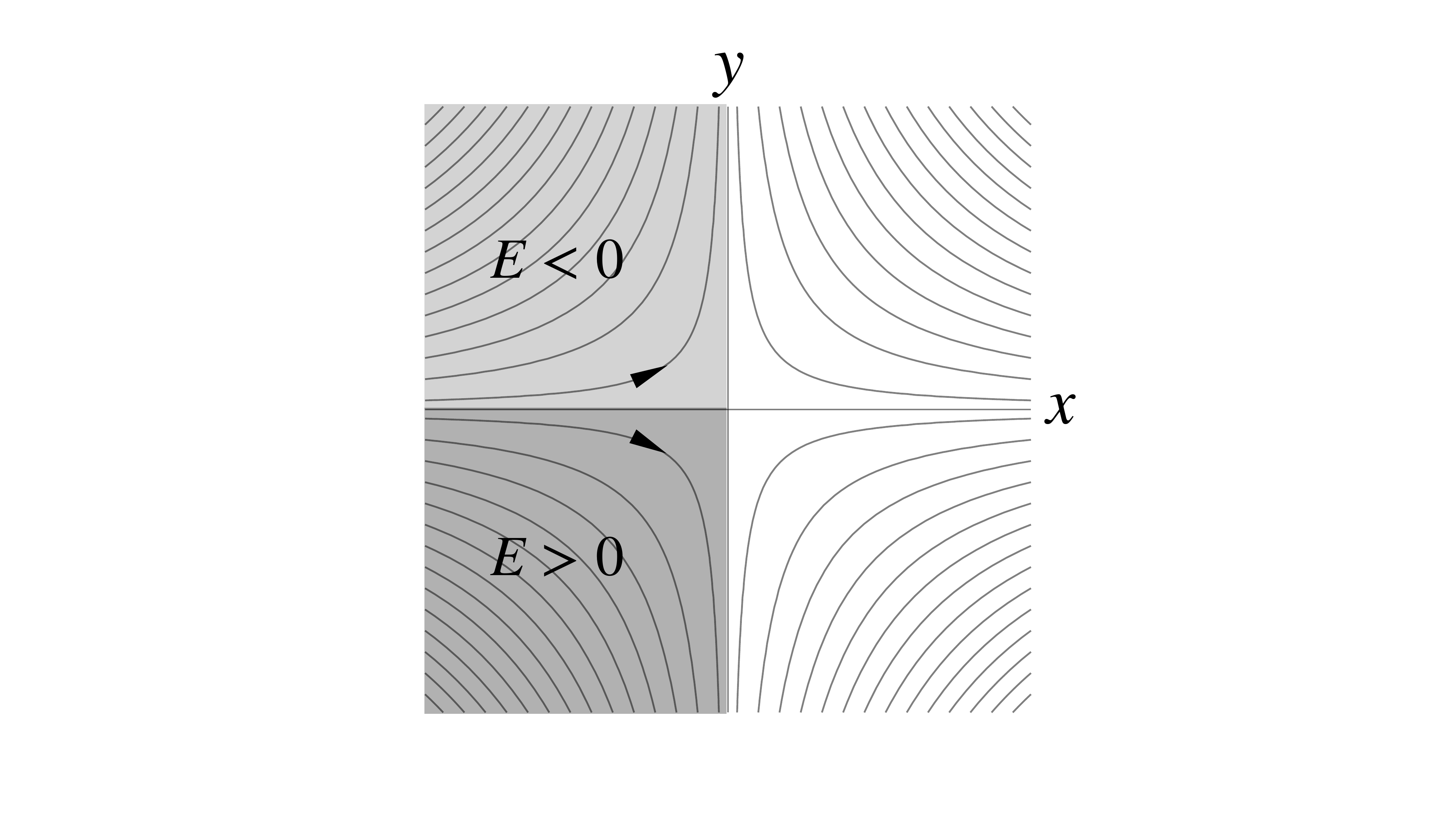}
\vspace*{-1cm}
\caption{Contours of the potential 
$V = \lambda xy$ in the plane of a 2d electron gas 
confined to the shaded regions in a uniform magnetic field $B$ perpendicular
to the plane. $E=qV$ is the electric potential energy, and  $q\lambda>0$ is assumed.
The chiral edge modes localized at $x=0$ propagate at speed
$\frac{\lambda}{B} y$, with a ``horizon'' at $y=0$, 
and exhibit Hawking radiation at the temperature 
$T_H = \frac{\hbar}{2\pi} \frac{\lambda}{B}$.}
\label{fig:Hawking}
\end{figure}

Stone's model \cite{Stone:2012cx} consists of a 2d electron gas 
confined to the half-plane $x<0$,
filling the lowest Landau level (LLL) 
in a uniform magnetic field $B$  
normal to the plane and a quadrupolar electrostatic potential 
\be\label{V}
V = \lambda xy
\ee
with $\lambda > 0$, where $x$ and $y$ are Cartesian coordinates
(see Fig.~\ref{fig:Hawking}).
%
Since the gas is the filled LLL, 
there are no low lying bulk degrees of freedom, 
but there are gapless chiral edge modes. 
A possible experimental realization of this system
is discussed in Section 5 of the published version of 
Ref.~\cite{Stone:2012cx}.

The classical guiding center drift velocity, 
to be derived below, is given by 
\be\label{speeds}
v^x = -\frac{\lambda}{B}\,x, \quad v^y = \frac{\lambda}{B}\, y\,.
\ee
Taking $\lambda/B$ to be positive, 
all electrons in the $x<0$ half-plane have $v^x>0$, 
and those with $y>0$ have $v^y >0$ while those with $y<0$ have $v^y<0$. 
In both cases they approach the $x=0$ edge where the drift velocity 
has only a nonzero $y$ component. 
The quantized edge modes are described by a massless chiral field
theory in an
effective $(1+1)$-dimensional edge spacetime at $x=0$.
The edge mode speed $v^y$ \eqref{speeds} suffices
to determine only the conformal metric, but one can read off the effective
metric by comparing the Hall effect edge mode Hamiltonian \cite{Stone:1990iw} 
to that for a minimally coupled chiral fermion in $(1+1)$-dimensional spacetime \cite{Sanielevici:1987zn}.\footnote{The expression for the Hamiltonian in the cited papers is correct only for the case of constant $v$, but the derivation
method of Ref.~\cite{Sanielevici:1987zn} extends to the non-constant case. 
(Ref.~\cite{Stone:1990iw} also 
discusses a bosonic description of the edge modes as small amplitude deformations of the boundary of the LLL droplet.)}
This effective metric is given by 
\be\label{2dmetric}
ds^2 = (v^y)^2 dt^2 - dy^2,
\ee
where $t$ is the laboratory time coordinate.
Since $v^y$ vanishes as $y$ goes to zero, 
this corresponds to a black hole spacetime
with horizon at $y=0$ (but see below for another interpretation).
In the black hole analogy, as described in \cite {Stone:2012cx},
$y<0$ is the ``inside'' and $y>0$ is the ``outside''.
We shall adopt this language, although the situation is
symmetric with respect to reflection $y\rightarrow - y$,
apart from a change of sign of the energy.

Consider electrons that come in from $x\rightarrow -\infty$.
We denote their charge by $q$, and assume that 
$q\lambda > 0$, hence for electrons $\lambda$ and $B$ are 
assumed negative.\footnote{The results in the paper depend
only on the signs of $\lambda/B$ and $q\lambda$, so the
same results are obtained if $q$, $\lambda$ and $B$ are 
all positive rather than negative.}
Classically, an electron 
with $y<0$ has positive energy $E=qV$ 
(where $q$ is the particle charge, and we measure energy 
relative to the constant kinetic energy of the gyro motion
in a given Landau level) and
always falls deep into the black hole, toward $y\rightarrow -\infty$.  
In contrast, an electron with $y>0$
has negative energy and moves away from the black hole,
toward $y\rightarrow +\infty$.
Quantum mechanically, in the microscopic theory \cite{Stone:2012cx},
a positive energy electron 
wavefunction extends from the third quadrant into the second quadrant, where classically
only negative energy states exist. As it propagates toward increasing $x$ it splits into 
components that travel up and down the $y$ axis, and spills partially
into the positive $x$ side (even though classically a particle that starts with 
negative $x$ would always remain on the negative $x$ half-plane).
Similarly, a negative energy electron wavefunction 
extends from the second quadrant into the third quadrant where classically
only positive energy states exist, 
and it too propagates both up and down, straddling the $y$ axis, and spilling
partially to positive $x$.
An observer sitting at the edge at $y=+\infty$ therefore 
sees a flux of positive energy 
electrons  and positive energy holes in the otherwise filled 
negative energy LLL sea.

The similarity to Hawking radiation is more than superficial:
using both single particle and second quantized many particle
analyses, Stone \cite{Stone:2012cx} found that
the flux of these edge mode excitations is thermal, with the Hawking temperature 
\be\label{TH}
T = \frac{\hbar \kappa}{2\pi}\,, \qquad \mbox{with}\qquad  \kappa = \left.\frac{dv^y}{dy}\right|_{y=0} = \frac{\lambda}{B}\,.
\ee
Here $\kappa$ is the surface gravity of the horizon at 
$y=0$ in the $(1+1)$-dimensional spacetime with 
metric \eqref{2dmetric}, 
defined with respect to the horizon-generating 
Killing vector field $\partial_t$; that is, $\kappa$ is defined with respect
to the energy conjugate to laboratory time translation.

While the metric \eqref{2dmetric} looks qualitatively like that of 
the static region outside a black hole horizon, 
it is actually {\it flat spacetime in hyperbolic polar (Rindler) coordinates}, 
since $v^y\propto y$. The hyperbolic (boost) angle $\eta$ is proportional to the
laboratory time, $\eta = \kappa t$. We shall refer to this effective flat spacetime
as the analog Minkowski spacetime.\footnote{Since the laboratory time goes to infinity at the horizons, 
only the right and left Rindler wedges are present in the analog spacetime,
while the future and past wedges --- the black and white hole interiors if we think
of it as a black hole --- are absent.}
The worldline of a laboratory observer at fixed $y$ is accelerated 
with respect to the effective metric; so if, as we shall argue at the end of section 
\ref{secHawking}, the edge mode
field is in the analog Minkowski vacuum state, a laboratory detector would,
according to the Unruh effect, perceive a thermal bath at the temperature
\eqref{TH}. Note that 
although the proper acceleration (with respect to the effective metric)
of the laboratory detector goes to zero as $y$ goes to infinity, 
the detected lab temperature remains constant, since lab time corresponds to the
constant $\kappa$ times the hyperbolic angle, and {\it not} to the proper time
along the hyperbola. 
Finally, there is one more reason why the analog Unruh effect
here resembles the Hawking effect: since the edge
modes are chiral, only radiation {\it outgoing} from the horizon exists.

\section{Guiding center effective theory}

In the guiding center approximation  (GCA), the gyro motion of the 
charges is unresolved, 
and one follows only the motion of the center of the gyro orbit,
i.e., the so-called ``guiding center'' 
motion \cite{alfven1950cosmical, northrop1963adiabatic}. 
Here we consider 
the motion of a charged particle confined to a two-dimensional plane, 
in a constant
magnetic field normal to the plane, and with an electrostatic potential $V$.
In the Lagrangian of the effective
classical theory, the contribution to the kinetic energy from the component
of velocity perpendicular to the magnetic field---which is the only contribution
in our two-dimensional setting---is 
replaced by the magnetic dipole energy
$\mu B$, and treated as a potential energy.\footnote{The 
energy of a magnetic dipole in a magnetic field 
is in general $-\vec \mu\cdot\vec B$. 
For gyro motion $\vec\mu$ is antiparallel to 
$\vec B$, for either sign of the charge $q$, so $\vec \mu\cdot\vec B<0$ always,
hence the magnetic dipole energy is $|\vec\mu||\vec B|=:\mu |B|$.}
The magnetic moment $\mu$ associated with the 
gyro motion is proportional to the angular momentum about the field line, and is an 
adiabatic invariant, so it is approximately conserved provided that the 
magnetic field and electrostatic potential are
nearly constant over the distance the particle travels during a gyro orbit. 
It is taken as  constant in the guiding center approximation. 
The  drift velocity contribution to the kinetic energy is subleading, 
and therefore neglected.  The two-dimensional case of guiding center motion was 
discussed in \cite{Witten:1978rg}, and quantum aspects of this system have been discussed 
in\cite{chan2016quantum, Chan:2017dex,
Klauder:1996rj}. 
 
The action can be found by coarse-graining the microscopic one
(see \cite{Jacobson:2024aap} for a pedagogical treatment). 
In two dimensions this reduces to 
\be\label{macroS}
S = \int\!dt \left(q\, A_i\,\dot x^i - \mu |B| - qV\right)\,,
\ee
where $q$ is the particle charge and $A_i$ is the magnetic vector potential
and $x^i$ represents the guiding center of the gyro motion, and $V$ is the 
electrostatic potential \eqref{V}.
The magnetic field strength $B$ is related to the vector 
potential via $\partial_i A_j - \partial_j A_i = B \epsilon_{ij}$,
where $\epsilon_{ij}$ is the area 2-form of the plane $\epsilon = dx\wedge dy$.\footnote{The action \eqref{macroS}, and thus the dynamics, 
is invariant under simultaneous sign change of $q$, $B$, and $\lambda$.}
Hamilton's principle for the action \eqref{macroS} 
with constant $B$
implies the 
equation of motion
\begin{equation}
    B\epsilon_{ij} \dot x^j = \partial_i V
\end{equation}
whose Cartesian components 
yield the drift velocities $v^x$ and $v^y$ \eqref{speeds}.

To canonically quantize this theory we start by evaluating the conjugate momenta,
\be
p_i := \frac{\partial L}{\partial \dot x^i} = qA_i(x)\,.
\ee
Since this defines the momenta as functions of the coordinates, it actually
corresponds to a pair of primary constraints.\footnote{The 
theory of constrained Hamiltonian systems was first 
worked out by Dirac and Bergmann 
(see \cite{Brown:2022iez} for a concise summary and references).}
Instead of working with
phase space coordinatized by $(x^i,\,p_j)$, one may pass to 
the reduced phase space in which the constraints are 
imposed ab initio. One restricts to the constraint surface
$p_i =  qA_i(x)$, which is coordinatized by $x^i$ alone, 
and the original symplectic form $dp_i\wedge dx^i$ is replaced by 
\be\label{omega}
\omega = d(qA_i)\wedge dx^i = q\partial_j A_i dx^j\wedge dx^i 
=: \frac{qB}{2} \epsilon_{ij}\, dx^i\wedge dx^j\,.
\ee
The corresponding Poisson bracket is called the {\it Dirac bracket},
which we denote by $\{\,,\,\}$.
The Dirac bracket
between functions $f(x)$ and $g(x)$ is given by
\be
\{ f, g\} = - \frac{1}{qB} \epsilon^{ij} \frac{\partial f}{\partial x^i} \frac{\partial g}{\partial x^j}
\ee
where $\epsilon^{ij}$ is the inverse defined by $\epsilon^{ij}\epsilon_{kj}=\delta^i{}_k$.
In particular, the Dirac bracket of the Cartesian coordinates $x$ and $y$
is 
\be\label{calg}
\{ x, y\} = - \frac{1}{qB}\,.
\ee
The position 
coordinates thus do not Poisson commute with each other. Note that at this 
stage $x$ and $y$ can be any area-adapted coordinates
(i.e.\ with $\epsilon_{xy}=1$). They need not be Cartesian,
since the Euclidean metric of space does not appear in the action.

In this reduced phase space formalism the Hamiltonian is
\be\label{H}
H = \mu\, |B| + qV\,.
\ee
With constant $B$ the brackets \eqref{calg} may be canonically quantized as 
\be\label{constBCC}
[ \wh x, \wh y ] = - i\ell^2\,,
\ee
where 
\be
\ell=\sqrt{\frac{\hbar}{qB}}
\ee
is the {\it magnetic length}.
The quantized theory thus lives on a noncommutative plane,
with non-commutativity scale equal to $\ell$. 
In the microscopic theory this length  sets the scale of a 
circular lowest Landau level wavefunction; hence, even though the 
gyro motion is not resolved in the effective theory, the length 
scale of the smallest quantum gyro motion governs the 
effective theory via the commutation relation \eqref{constBCC}.

According to the Stone-von Neumann theorem, the only (regular) unitary irreducible 
representation of the Heisenberg algebra \eqref{constBCC} that exponentiates 
to a representation of the Heisenberg group is the familiar one on $L^2(\bb R)$,  
carried by complex-valued squared-integrable wavefunctions of either variable. 
In the $x$ representation we have
\ba\label{xyrep}
\wh x \psi(x) &= x \psi(x) \\
\wh y \psi(x) &= i\ell^2 \frac{\partial\psi}{\partial x}(x)\,. \label{x2rep}
\ea
Note that, although the representation is unique, the quantization is ambiguous since there is no
preferred class of area-adapted coordinates. Reaching beyond the effective theory, 
we appeal to the Euclidean metric that 
appears in the action of the microscopic theory, and thus specialize to Cartesian coordinates.

\section{Guiding center quantum Hall and Hawking effects}
\label{sec:HH}

Let us now first analyze the quantum Hall effect in a linear potential, and then
turn to Stone's Hall effect model of Hawking radiation \cite{Stone:2012cx}.
Since we restrict to the case of constant $B$,
$\mu B$ is just an additive constant in the Hamiltonian that affects
only the global phase evolution of the wave function, so for notational brevity
we shall drop this term. We thus measure all energies relative to 
$\mu B$.

\subsection{Quantum Hall effect}

Consider a charged particle in two dimensions, 
in a uniform magnetic field 
normal to the plane and electrostatic potential
\be
V = -E^y y\,,
\ee
where $(x,y)$ form a pair of area-adapted coordinates. 
If we quantize these coordinates, the Heisenberg $\wh y$ operator
is constant and $d\wh x/dt = \frac{q}{i\hbar}[x,V] = E^y/B$, so the charge has 
a (quantum) drift velocity $v_{\rm drift} = E^y/B$.
In the Schr\"odinger picture, 
the Hamiltonian in the $x$-representation
is 
\be
\wh H = -i\hbar \frac{E^y}{B}\frac{\partial}{\partial x}\,.
\ee
The Schr\"odinger equation follows from the Schr\"odinger action 
$\int dt\,dx\, (i\hbar \psi^*\partial_t\psi - \psi^*\wh H\psi)$, whose invariance
under constant phase transformations $\psi\rightarrow e^{i\theta}\psi$ implies
that, when the Schr\"odinger equation is satisfied, the probability current 
$(J^t,J^x) = (\psi^*\psi, \frac{E^y}{B}\psi^*\psi)$ is conserved. 

With many electrons, the charge current density is 
$j^x = q\rho_n v_{\rm drift}$, where $q$ is 
the particle charge 
and $\rho_n$ is the particle number density.
If all of the states with the given value of the magnetic moment
$\mu$ are filled, $\rho_n$ is the same as 
for a filled Landau level in the microscopic theory,
namely $\rho_n=1/2\pi \ell^2$ \cite{Landau:1991wop}. 
The argument is parallel to the one used 
in \cite{Landau:1991wop}.
Consider a rectangular region with coordinate lengths $a$ and $b$ in the $x$ and $y$ directions. (Since the coordinate operators do not commute, this is meaningful only for lengths much longer than $\ell$).  
If we impose periodic boundary conditions in the $x$ direction, the allowed Fourier wavenumbers for $\psi(x)$  are $k_n=2\pi n/a$, with integer $n$, whose spacing is $2\pi/a$. 
These wavenumbers are equal to eigenvalues of $\wh y/\ell^2$
(according to the $x$-representation of the $\wh y$ operator \eqref{x2rep}) 
which in the rectangular region range from $0$ to 
$b/\ell^2$. The number of wavenumbers that fit in this interval is 
$(b/\ell^2)/(2\pi/a) = ab/2\pi\ell^2$, 
in agreement with the density of states for a Landau level in the microscopic theory.
The current density is therefore
$j^x = q^2E^y/2\pi\hbar$, and the transverse Hall resistivity is 
$\rho_{xy} := E^y/j^x = 2\pi\hbar/q^2$. We thus recover the
standard result for the quantum Hall resistivity.

\subsection{Hawking effect} 
\label{secHawking}
We now analyze Stone's black hole analog from the perspective of 
the quantized effective guiding-center theory.

The classical Hamiltonian (neglecting the constant $\mu B$ terms) is 
$H = q\lambda xy$,
which we quantize as
\be\label{Vhat}
\wh H =\frac{q\lambda}{2}\left( \wh x \wh y + \wh y \wh x\right)\,.
\ee
The solutions to the corresponding Heisenberg equations of motion 
are 
\be
\wh x(t) = e^{-\kappa t}\wh x(0)\qquad \mbox{and}\qquad\wh y(t) = e^{\kappa t}\wh y(0)\,,
\ee
with $\kappa=\lambda/B$ \eqref{TH}, 
which amount to a squeezing in the $x$ direction and inverse stretching in the $y$ 
direction. In fact, the time evolution operator for the Hamiltonian \eqref{Vhat}
is precisely a squeeze operator.

For the Schr\"odinger picture, 
in the $x$-basis we have $\wh y = i\ell^2\frac{d}{dx}$ \eqref{xyrep}, and in the 
$y$-basis $\wh x = -i\ell^2\frac{d}{dy}$.
In the $x$-basis the Schrodinger equation reads
\be\label{StoneSE}
\frac{\partial\psi}{\partial t} - \kappa x \frac{\partial\psi}{\partial x}  - \frac{\kappa}{2} \psi = 0\,,
\ee
which can be integrated  to obtain
\be\label{psi(t,x)1}
\psi(t,x) = f\left(x e^{\kappa t}\right) e^{{\kappa t}/{2}}\,,
\ee
where $f$ is an arbitrary square-integrable complex function.
Note that if $\psi(0, x) = f(x)$ is initially supported  in the $x<0$ region, it will evolve towards $x=0$, but never cross into the $x>0$ region.
In the $y$-basis the Schr\"odinger equation reads
\be\label{StoneSEy5}
\frac{\partial\wt\psi}{\partial t} + \kappa y \frac{\partial\wt\psi}{\partial y} + \frac{\kappa}{2}  \wt\psi = 0\,,
\ee
for which 
the solution has the form 
\be\label{psi(t,y)1}
\wt\psi(t,y) = g\left(y e^{-\kappa t}\right) e^{{-\kappa t}/{2}}\,.
\ee
This demonstrates that if the particle starts with a wavefunction supported in 
either $y>0$ or $y<0$ it will remain in the same region.

It might therefore seem
at first that there will be no Hawking radiation in the effective theory.
However,   
since the electrons are assumed 
to enter from $x = -\infty$ their
wavefunction is necessarily nonvanishing for both signs of $y$. This is because
the $y$ representation is related to the $x$ representation by
Fourier transform, and
the  Fourier transform of a (sufficiently regular, non-zero) function $\psi(x)$ supported only on $x<0$ cannot vanish on any open set (since $\int_{-\infty}^0 dx\,  e^{ixy/\ell^2} \psi(x)$ is analytic in the lower-half complex $y$-plane). 
The portion of the wavefunction in $y>0$ will be pushed upwards (away from the black hole) while the portion in $y<0$ will keep falling down into the black hole. 
Hawking radiation thus arises from 
the condition that the incoming particle states are confined to the $x<0$ half 
of the non-commutative $xy$ plane.

The Hawking radiation rate depends on the particle energy.  To evaluate it 
we need to compute the probability currents in the positive and negative $y$ directions,
given an energy eigenstate corresponding to a particle that propagates in from
negative $x$. 
The Hamiltonian eigenvalue equation for $\wt\psi(y)$
can be expressed, using the commutation relation \eqref{constBCC} and
the $y$-representation expression $\wh x = -i\ell^2 \frac{\partial}{\partial y}$, as
\be\label{1stode}
y\frac{\partial\wt\psi}{\partial y} = \left( -\frac{1}{2} + i\Delta_E\right) \wt\psi\,,
\ee
where
\be
\Delta_E := \frac{E}{\hbar\kappa} \, .
\ee
The solutions are
\be\label{wtpsiE}
\wt\psi(y) = \left\{\begin{matrix}
\alpha_+\, y^{-\frac{1}{2} + i\Delta_E} \,,\quad y>0\\
\alpha_-\, (-y)^{-\frac{1}{2} + i\Delta_E} \,,\quad y<0 
\end{matrix}\right.
\ee
for complex constants $\alpha_+$ and $\alpha_-$.\footnote{That there is a two parameter
family of solutions, despite equation \eqref{1stode} being of first order, 
is a consequence of the singular nature of the highest derivative term at $y=0$.}
The relation between these constants is fixed by the 
physical requirement that the particle entered the system from 
$x<0$, and therefore (according to the solution \eqref{psi(t,x)1})
$\psi(x)$ has vanishing amplitude at $x>0$. 
The corresponding wave function in the $y$ representation is expressed in
terms of that in the $x$ representation via the integral transform
\be
\la y|\psi\ra = \int dx\, \la y|x\ra \la x|\psi\ra\,.
\ee
Since $\wh x = -i\ell^2 \frac{d}{dy}$, the $x$ eigenfunction in the 
$y$ representation is given by $\la y|x\ra\propto e^{ixy/\ell^2}$.
The wavefunctions $\la y|\psi\ra$ and $\la x|\psi\ra$
are thus related by Fourier transform.
It follows that, for electrons confined to $x<0$,  
$\la y|\psi\ra$ is analytic in 
the lower-half complex $y$ plane.
We may therefore obtain the value of $\la -y|\psi\ra$, for positive $y$, from the value
of $\la y|\psi\ra$  by the replacement $y\rightarrow e^{-i\pi}|y|$,
which yields the relation\footnote{This method is identical to that
introduced by Unruh \cite{Unruh:1976db} when 
finding the relation between modes of positive free-fall
frequency and definite Killing frequency when deriving the 
Unruh and Hawking effects.} 
\be\label{ratio1}
\alpha_- = i e^{\pi\Delta_E} \alpha_+\,.
\ee
With this relation we can compute the relative probabilities for the
particle to propagate to positive or negative $y$ values, using the 
probability current. 

From global phase invariance of the Schr\"odinger action 
$\int dt\,dx\,(i\hbar \psi^*\partial_t\psi - \psi^*\wh H\psi)$
(or directly from the Schr\"odinger equation \eqref{StoneSEy5}) 
we obtain the continuity equation 
\be
\frac{\partial}{\partial t} |\wt\psi|^2 + \frac{\partial}{\partial y}\left(\kappa y |\wt\psi|^2\right) = 0\,,
\ee
which identifies the $y$-component of the probability 
current as
\be
\wt J^y = \kappa y |\wt\psi(y)|^2\,.
\ee
Although the eigenfunctions are not normalizable, the probability current is well-defined,
\be\label{Jy}
\wt J^y = \left\{\begin{matrix}
\kappa|\wt\psi_+|^2 \,,\quad y>0\\
-\kappa|\wt\psi_-|^2 \,,\quad y<0 \,.
\end{matrix}\right.
\ee
For $y>0$ the particles move up, while for $y<0$ they move down, compatible with the classical behavior. 
The source discontinuity at $y=0$ arises because we are working with nonnormalizable states. 
For a normalizable wavefunction, the current is everywhere conserved, and 
this source can be understood as follows. Looking at the 
solution \eqref{psi(t,y)1} for $\wt\psi(t,y)$, 
we see that in the far past the wavefunction is squeezed in a region arbitrarily close to the origin; as time passes forward it spreads out away from the origin, compatible with \eqref{Jy}. 
On the other hand, in the $x$-basis, the wavefunction $\psi(t,x)$ is spread across the negative $x$-axis in the far past, and as time passes forward it accumulates closer and closer to the origin. This suggests the heuristic picture of the probability current being ``fed'' from the $x$-axis, through the origin, into the $y$-axis. (This 
interpretation is consistent
with the microscopic description, in which the $x$ and $y$ coordinates of the
particle can be simultaneously specified.)

Given the amplitude ratio \eqref{ratio1}, we thus have
\be
\wt J^y \propto 
\left\{\begin{matrix}
\hspace{9mm} 1\,,\hfill y>0 \\
- e^{2\pi\Delta_E} \,,\quad y<0
\end{matrix}\right.
\ee
where the proportionality factor is positive and independent of $y$.
The probability for the observer at $y = +\infty$ to detect the particle 
is given by the fraction of the total conserved current,
\be\label{pE}
p_E = \frac{\wt J_{+\infty}}{-\wt J_{-\infty}+\wt J_{+\infty}} = \frac{1}{e^{\frac{2\pi}{\hbar \kappa}E }+1}\,.
\ee
This is a Fermi distribution with temperature
\be\label{TF}
T = \frac{\hbar\kappa}{2\pi} \,,
\ee
which is in agreement with Stone's analysis (his $\lambda$ is our $q\lambda$, and
$\kappa$ is defined in \eqref{TH}). 
As already mentioned in \eqref{TH}, the temperature \eqref{TF} 
is $\hbar/2\pi$ times the surface gravity of the analog black hole. 
If the state is the Fermi sea
incoming from $x<0$, with all states occupied, 
a flux of particles and holes characteristic of the Hawking effect results. 
The probability  
for each electron of positive energy $E>0$ to escape to $y>0$ 
is \eqref{pE}, and the
probability 
for each
electron of negative energy $-E<0$ {\it not} to escape, i.e., for a hole
of positive energy $E$ to be emitted, is the same,
since $1-(1+e^{-E/T})^{-1}
= (1+e^{E/T})^{-1}$. 

Using the microscopic theory, Stone also showed that 
the many-body state of the Fermi sea of one filled Landau level
flowing in from negative $x$,
$|\Omega_{\it in}\ra$,
takes the form of a thermofield double in terms of excitations 
of the ``out-vacuum" $|\Omega_{\it out}\ra$, 
whose restriction to either the positive or negative
$y$ factor of the Hilbert space 
is precisely a Gibbs ensemble. As shown in detail in the Appendix,
we find precisely the same structure 
using the guiding center theory:
\be\label{Oin}
|\Omega_\text{\it in}\ra \propto \exp\left[ i \int_0^\infty\!dE\, e^{-\beta E/2} \big( \wt a_-(E) \wt a_+^\dag(E) + \wt a_-^\dag(-E) \wt a_+(-E) \big)\right] |\Omega_\text{\it out}\ra\,.
\ee
Here $|\Omega_{\it out}\ra$ is the state with filled negative energy and empty positive energy states supported on $y>0$, and filled positive energy and empty negative energy states supported on $y<0$. The $\wt a_\pm$ and $\wt a^\dagger_\pm$ operators are the usual fermionic anticommuting 
annihilation and creation operators for the positive and negative $y$ factors of the out
Hilbert space ($i$ in the exponent could be absorbed by changing their phase).
Particles and holes are correlated in pairs: positive energy particles in $y>0$ with partner holes in $y<0$, and negative energy particles in $y<0$ with partner holes in $y>0$.
The form of the state is familiar from standard representations of the Minkowski vacuum in the Fock space of left and right Rindler wedge modes. That is, $|\Omega_{\it out}\ra$ corresponds to the 
product of left and right Rindler vacua 
(i.e., the ground states of the left and right analog boost Hamiltonians, 
 which generate respectively backward and forward laboratory time translation scaled by $\kappa$),
while $|\Omega_{\it in}\ra$ corresponds to
the analog Minkowski vacuum, justifying 
our assertion at the end of section \ref{sec:StoneModel}. 
All features of the Hawking-Unruh effect in the Hall edge mode model are
thus accurately predicted in the  quantized guiding center theory.

\section{Discussion}

In this paper we have explicated
Stone's analogue model of Hawking radiation 
by edge modes in a quantum Hall system with a 
quadrupolar electric field  \cite{Stone:2012cx}. We first clarified the 
identification of the effective metric for the chiral 
edge modes, which is a Minkowski metric, 
and noted that the laboratory time is proportional to 
the boost angle in the Rindler wedges. For this reason, 
and since the edge modes are chiral, the lab observer
measures a thermal flux of edge modes coming from the 
horizon, although the state is the vacuum 
with respect to the analog Minkowski 
space. This Hawking radiation is
thus strictly equivalent to the Unruh effect, but characterized by
a different clock: the boost angle rather than the proper time
for particular accelerated observers.

We next derived the physical consequences of the model from scratch,
starting with the guiding center effective theory, which is a
coarse-grained approximation in which the gyro motion of the 
charges is not resolved. The single particle phase space for this theory
is the spatial plane of the system, with canonically conjugate coordinates, 
so its quantization is the noncommutative plane. 
We found perfect agreement with the microscopic theory 
for the prediction of the Hawking radiation, 
even though the wavefunction depends only on the Cartesian coordinate 
$x$ or only on $y$. Due to the non-commutativity of the space, 
a quantum state characterized by a wave function
$\psi(x)$ can also be described by a function of $y$, which is
its Fourier transform ${{\cal F}}[\psi](y)\propto \int dx\,  e^{ixy/\ell^2}\psi(x)$. 
Since the particles come in from $x<0$ and $\psi(x)$ vanishes for $x>0$, 
${{\cal F}}[\psi](y)$ is analytic in the lower-half complex 
$y$ plane, hence necessarily has support at both positive and negative $y$. 
This yields the same current ratio as Stone found in the 
microscopic theory. Finally, we showed that in the guiding center many-body 
theory, with an incoming filled Fermi sea, 
the state has the thermofield double form, with particles on one side of 
the horizon correlated with holes on the other side, in agreement with what Stone 
found via second quantization of the microscopic theory. 

We conclude with some comments on the mechanism and the model.

\paragraph{Is tunneling the mechanism?}
A classical charge 
with a very small gyroradius drifts along an equipotential of the electric field, and as such its motion is confined to one quadrant of the plane. In the quadrupole field the equipotential curves come in pairs of equal potential branches, related by inversion through the origin of the plane. 
One might thus expect that quantum tunneling to the opposite branch accounts for the emission of holes and particles, and indeed this is the language used throughout Ref.~\cite{Stone:2012cx}. However it seems to us that such a tunneling is not a faithful description of the process.
It is true that in the microscopic theory the wave function for a particle entering from negative $x$ partially spills into positive $x$; nevertheless, as the particle propagates in the $y$ direction
it straddles the $y$ axis, no more on the positive $x$ side than on the negative $x$ side
at large values of $|y|$.
Also, in the guiding center theory, the same Hawking current is predicted, yet there is strictly zero amplitude for the particle to lie at a positive value of $x$. Instead the reason for the Hawking emission is that the wave function of a quantum particle 
confined to negative $x$
is not localized in one quadrant, at either positive or negative $y$ values, but rather overlaps both signs of 
$y$. Thus, long before any putative tunneling event might have taken place, 
an incoming wavepacket 
already possesses a nonzero amplitude to propagate to both positive and negative $y$ directions as it approaches $x=0$ from the negative $x$ side.

\paragraph{Why the Fermi-Dirac distribution?}

At the single particle level, the calculation of the current knows nothing about the 
statistical nature, Bose or Fermi, of the particle, so why does the Fermi distribution 
appear in the current ratio \eqref{pE}? As far as we can tell this is a fluke. 
In any case, 
at the many particle level we assumed Fermi statistics when interpreting the
absence of a negative energy electron as a positive energy hole in the paragraph under
Eq.~\eqref{TF}, and when defining the in and out states that 
are related via \eqref{Oin}.

\paragraph{What about the missing future and past wedges of the analog Minkowski space?}
As explained in section \ref{sec:StoneModel}, the analog Minkowski space  
for the chiral edge modes includes only the left and right Rindler wedges (in our coordinates, the 
positive and negative $y$ wedges), and is missing the
future and past wedges. Does this incompleteness of the analog spacetime correspond to
some sort of incompleteness of the effective $(1+1)$-dimensional field theoretic 
description of the edge modes? The edge modes are chiral, propagating (in the future direction) only to the right in the right wedge and only to the left in the left wedge. They cannot enter the future wedge from the past, so the missing future wedge does not correlate to missing physics. The missing past wedge is another matter, since
in the analog Minkowski spacetime chiral modes must
enter the right and left wedges from the past wedge. This geodesic incompleteness of
the analog spacetime has no direct effect in the laboratory, since the modes would emerge from 
the past horizon infinitely far in the past with respect to laboratory time.
Correspondingly, the quantized guiding center theory is complete on its own terms. 
On the other hand, in the edge spacetime description 
the outgoing modes all emerge from an infinite density of 
states at $y=0$, which is physically suspect, especially because it relies on localizing 
particles in a region arbitrarily small compared to the magnetic length $\ell$ that 
sets the noncommutativity scale. 
In fact, the incompleteness of the 
analog Minkowski space is a sign that the dynamics transcends the
edge spacetime description. Rather than emerging from a past wedge that does not exist, the outgoing particles on the edge actually come from the bulk half-plane.

\section*{Acknowledgements}
We thank Michael Stone for helpful discussions. 
This work was supported in part by NSF grant PHY-2309634
and by Perimeter Institute for Theoretical Physics. 
Research at Perimeter Institute is supported in part by the Government of Canada through the Department of Innovation, Science and Economic Development and by the Province of Ontario through the Ministry of Colleges and Universities.

\appendix
\section{Second quantization}

In this appendix we describe the Fock space representation of the many-body system, and construct the relationship between the ``in-vacuum'' (describing the actual quantum state of the system) and the ``out-vacuum'' (describing the vacuum state from the perspective of an asymptotic observer along the edge). 

Consider the basis of single particle eigenfunctions of energy in the $x$ representation,
\ba
\psi_-(x,E) &= \frac{\sqrt{\beta}}{2\pi} \left\{
\begin{array}{cl}
(-x)^{-1/2 - i\beta E/2\pi} & \text{for $x < 0$} \label{psi-xE}\\
0 & \text{for $x > 0$}
\end{array}\right. \\
\psi_+(x,E) &= \frac{\sqrt{\beta}}{2\pi} \left\{
\begin{array}{cl}
0 & \text{for $x < 0$} \\
x^{-1/2 - i\beta E/2\pi} & \text{for $x > 0$}
\end{array}\right. \label{psi+xE}
\ea
with $\beta := 1/T = 2\pi/\hbar\kappa$. These are normalized in the continuous sense,  
\be
\int\!dx\, \psi_\pm(x,E)^* \psi_{\pm'}(x,E') = \delta(E,E')\delta_{\pm\pm'}\,.
\ee
Let us introduce annihilation operators $a_\pm(E)$,  
such that $a_\pm^\dag(E)$ creates a particle in the state $\psi_\pm(E)$
above a Fock vacuum $|0\ra$, that is,
\be
\la x| a_\pm^\dag(E)|0\ra = \psi_\pm(x,E)\,.
\ee
The (fermionic) operators $a_\pm(E)$ satisfy, by definition, ladder algebra
commutation relations
\ba
\gb{a_\pm(E),a_{\pm'}^\dag(E')} &= \delta(E,E')\delta_{\pm\pm'} \\
\gb{a_\pm(E),a_{\pm'}(E')} &= 0\,,
\ea
where $\gb{\cdot,\cdot}$ denotes the graded bracket.\footnote{If $\scr O$ and $\scr O'$ have fermionic weights $s$ and $s'$, respectively, then $\gb{\scr O, \scr O'} = \scr O \scr O' - (-1)^{ss'} \scr O' \scr O$. (The fermionic weights of bosonic and fermionic operators are $0$ and $1$, respectively.) The graded bracket satisfies the graded product rule,
$\gb{\scr O, \scr O'\scr O''} = \gb{\scr O, \scr O'}\scr O'' + (-1)^{ss'} \scr O'\gb{\scr O, \scr O''}$.}
In the $y$ representation we have
\be\label{yapmdag}
\la y| a_\pm^\dag(E)|0\ra = \int\!dx\, \la y|x\ra \la x| a_\pm^\dag(E)|0\ra  = {\cal F}[\psi_\pm(x,E)](y)\,,
\ee
where ${\cal F}$ denotes the Fourier transform,
\be
{\cal F}[f(x)](y) = \frac{1}{\sqrt{2\pi\ell^2}}\int dx\, e^{ixy/\ell^2} f(x)\,.
\ee
A basis of eigenfunctions of energy in the $y$ representation, 
defined analogously to \eqref{psi-xE} and \eqref{psi+xE}, is given by
\ba
\wt\psi_-(y,E) &= \frac{\sqrt{\beta}}{2\pi} \left\{
\begin{array}{cl}
(-y)^{-1/2 + i\beta E/2\pi} & \text{for $y < 0$} \\
0 & \text{for $y > 0$}
\end{array}\right. \\
\wt\psi_+(y,E) &= \frac{\sqrt{\beta}}{2\pi} \left\{
\begin{array}{cl}
0 & \text{for $y < 0$} \\
y^{-1/2 + i\beta E/2\pi} & \text{for $y > 0$}\,,
\end{array}\right.
\ea
and we similarly introduce the corresponding annihilation operators 
$\wt a_\pm(E)$, satisfying ladder algebra commutation relations. 

The Fourier transform of 
$\psi_+(x,E)$ can be written as a linear combination of $\wt\psi_\pm(y,E)$,
\be
{\cal F}[\psi_+(x,E)](y) = \gamma_- \wt\psi_-(y,E) + \gamma_+ \wt\psi_+(y,E)\,.
\ee
Consequently, we obtain from \eqref{yapmdag} that
\be
\la y| a_+^\dag(E)|0\ra = \gamma_- \wt\psi_-(y,E) + \gamma_+ \wt\psi_+(y,E) = \gamma_- \la y| \wt a_-^\dag(E)|0\ra + \gamma_+ \la y| \wt a_+^\dag(E)|0\ra\,,
\ee
which (since the $a^\dagger$ and $\wt a^\dagger$ operators all create only single particle
states when acting on $|0\ra$) yields the operator relation
\be
a_+^\dag(E) = \gamma_- \wt a_-^\dag(E) + \gamma_+  \wt a_+^\dag(E)\,.
\ee
To compute the coefficients $\gamma$, we use the same analyticity trick employed in Sec.~\ref{secHawking}. Defining a variable $s= xy$, we have for positive $y$ 
\be
{{\cal F}}[\psi_+(x,E)](y) \propto \left(\int_0^\infty\!ds\, e^{is}s^{-1/2-i\beta E/2\pi}\right) y^{-1/2 +i\beta E/2\pi}\,,
\ee
hence  
\be
{{\cal F}}[\psi_+(x,E)](y) \propto  y^{-1/2 +i\beta E/2\pi}\,.
\ee
For positive $x$ the integral ${{\cal F}}[\psi_+(x,E)](y)$ 
is analytic in the upper-half (complexified) 
$y$-plane, so we have with the {\it same} proportionality factor
\be
{{\cal F}}[\psi_+(x,E)](-|y|) \propto  (e^{i\pi}|y|)^{-1/2 +i\beta E/2\pi}
= e^{-i\pi/2}e^{-\beta E/2} |y|^{-1/2 +i\beta E/2\pi}\,,
\ee
from which it follows that
\be
{{\cal F}}[\psi_+(x,E)](y) \propto  \wt\psi_+(y,E) -i e^{-\beta E/2} \wt\psi_-(y,E)\,.
\ee
and thus
\be
a_+^\dag(E) \propto \wt a_+^\dag(E) -i e^{-\beta E/2} \wt a_-^\dag(E)\,.
\ee
For the Fourier transform of
$\psi_-$ the calculation is nearly identical, but in this case the result is analytic in the
{\it lower}-half $y$-plane, so $e^{i\pi}|y|$ is replaced by $e^{-i\pi}|y|$, with the result that 
the coefficient of $\wt\psi_-$ relative to that of $\wt\psi_+$ is $e^{+i\pi/2}e^{+\beta E/2}$.
We thus have the relations
\be
a_\pm(E) \propto \wt a_+(E) \pm i e^{\mp\beta E/2} \wt a_-(E)\,.
\ee

The ``in-vacuum'' is defined by
\ba
a_-^\dag(E)|\Omega_\text{\it in}\ra &= 0 \label{invacm} \\
a_+(E)|\Omega_\text{\it in}\ra &= 0 \,,\label{invacp}
\ea
that is, in the $x$-representation, all the states in $x<0$ are filled 
and all the states in $x>0$ are empty.
The ``out-vacuum'' is defined by the conditions
\ba
\wt a_-(E) |\Omega_\text{\it out}\ra = 0 & \quad\text{for $E<0$} \\
\wt a_-^\dag(E) |\Omega_\text{\it out}\ra = 0 & \quad\text{for $E>0$} \\
\wt a_+(E) |\Omega_\text{\it out}\ra = 0 & \quad\text{for $E>0$} \\
\wt a_+^\dag(E) |\Omega_\text{\it out}\ra = 0 & \quad\text{for $E<0$}\,.
\ea
The first two lines say that at $y<0$ negative energy states are empty 
and positive energy states are filled; the last two lines say that at $y>0$
positive energy states are empty and negative energy states are filled. 

To find the relation between the in and out vacua, it is convenient to 
define $b_\pm$ operators
\ba
b_-(E) &=  \left\{
\begin{array}{cl}
\wt a_-(E) & \text{for $E < 0$} \\
\wt a_-^\dag(E) & \text{for $E > 0$}
\end{array}\right. \\
b_+(E) &=  \left\{
\begin{array}{cl}
\wt a_+^\dag(E) & \text{for $E < 0$} \\
\wt a_+(E) & \text{for $E > 0$}\,.
\end{array}\right.
\ea
The out-vacuum is thus defined by $b_\pm|\Omega_\text{\it out}\ra=0$. 
In this way, $b^\dagger$ describes either creation of a particle from an empty
state, or creation of a ``hole'' from a filled state. 
Moreover, the $b$ operators satisfy ladder algebra commutation relations for all $E$,
\ba
\gb{b_\pm(E),b_{\pm'}^\dag(E')} &= \delta(E,E')\delta_{\pm\pm'} \\
\gb{b_\pm(E),b_{\pm'}(E')} &=  0
\ea
The in-vacuum can thus be expressed as
\be
|\Omega_\text{\it in}\ra = F(b_-^\dag, b_+^\dag)|\Omega_\text{\it out}\ra
\ee
for some functional $F$ of the creation operators. 
The conditions defining the in-vacuum, \eqref{invacm} and \eqref{invacp}, 
can be written 
compactly as 
\be\label{compact}
\left(b_\pm(E) \pm i e^{-\beta |E|/2} b_\mp^\dag(E) \right) F |\Omega_\text{\it out}\ra = 0 
\ee
for all $E$. 
Since $b_\pm(E)$ annihilates $|\Omega_\text{\it out}\ra$ we have 
\be
b_\pm(E) F |\Omega_\text{\it out}\ra = \gb{b_\pm(E), F} |\Omega_\text{\it out}\ra
\ee
and, as the resulting operators acting on $|\Omega_\text{\it out}\ra$ 
in \eqref{compact}
contain only creation operators, the following operator equations must hold:
\be\label{vaceqn}
\gb{b_\pm(E),F} \pm i e^{-\beta |E|/2} b_\mp^\dag(E) F = 0\,.
\ee
The bracket with $b_\pm(E)$ amounts to a (fermionic) functional derivative
with respect to $b_\pm^\dag(E)$, so one readily guesses that a solution
is given by 
\be
F_0 := \exp\left[i \int\!dE\, e^{-\beta |E|/2} b_-^\dag(E) b_+^\dag(E) \right]\,,
\ee
as is easily verified using the graded product rule. It remains only to show that
$F_0$ is the unique solution, up to a multiplicative constant.
To establish uniqueness, 
we suppose $F$ is also a solution to \eqref{vaceqn}, and consider the bracket
\be\label{bF0F}
\gb{b_\pm(E),F_0^{-1}F}\,.
\ee
Since $F_0^{-1}F$ involves only $b^\dagger$ operators, and the bracket satisfies the (graded)
product rule, vanishing of \eqref{bF0F} would imply that $F_0^{-1}F=c$ for some constant scalar
$c$, so that $F = c F_0$.
Indeed, we have
\ba
\gb{b_\pm(E),F_0^{-1}F} &= \gb{b_\pm(E),F_0^{-1}}F + F_0^{-1}\gb{b_\pm(E),F}\nonumber\\
&= i e^{-\beta |E|/2} b_\mp^\dag(E) F_0^{-1}F + F_0^{-1} (-i e^{-\beta |E|/2} b_\mp^\dag(E) F)\nonumber\\
&=0\,,
\ea
where we used that $F_0^{-1}$ is bosonic.
Finally, in terms of the original particle operators the state $F_0|\Omega_{\it out}\ra$
takes the form 
\be
|\Omega_\text{\it in}\ra \propto \exp\left[ i \int_0^\infty\!dE\, e^{-\beta E/2} \big( \wt a_-(E) \wt a_+^\dag(E) + \wt a_-^\dag(-E) \wt a_+(-E) \big)\right] |\Omega_\text{\it out}\ra\,.
\ee

\bibliographystyle{JHEPmod}
\bibliography{HallHawking}

\end{document}